\documentclass[prb,nofootinbib,twocolumn,superscriptaddress]{revtex4} 

\usepackage{graphicx}
\usepackage{dcolumn}
\usepackage{bm}
\usepackage{threeparttable}
\usepackage{times}
\usepackage{mathptmx}
\usepackage{lscape}
\usepackage{natbib}
\usepackage{amsmath}
\usepackage{amssymb}
\usepackage{braket}
\usepackage{comment}
\usepackage{color}

\def\degree{\kern-.2em\r{}\kern-.3em}

\begin{document}


\title{ Landscape of Configurational Density of States for Discrete Large Systems  }

\author{Koretaka Yuge}
\affiliation{
Department of Materials Science and Engineering,  Kyoto University, Sakyo, Kyoto 606-8501, Japan\\
}%

\author{Tetsuya Taikei}
\affiliation{
Department of Materials Science and Engineering,  Kyoto University, Sakyo, Kyoto 606-8501, Japan\\
}%

\author{Kazuhito Takeuchi}
\affiliation{
Department of Materials Science and Engineering,  Kyoto University, Sakyo, Kyoto 606-8501, Japan\\
}%

\begin{abstract}
{ For classical many-body systems, our recent study reveals that  expectation value of internal energy, structure, and free energy can be well characterized by a single specially-selected microscopic structure. This finding relies on the fact that configurational density of states (CDOS) for typical classical system before applying interatomic interaction can be well characterized by multidimensional gaussian distribution. Although gaussian distribution is an well-known and widely-used function in diverse fields, it is quantitatively unclear why the CDOS takes gaussian when system size gets large, even for projected CDOS onto a single chosen coordination. Here we demonstrate that for equiatomic binary system, one-dimensional CDOS along coordination of pair correlation can be reasonably described by gaussian distribution under an appropriate condition, whose deviation from real CDOS mainly reflects the existence of triplet closed link consisting of the pair figure considered. The present result thus significantly makes advance in analytic determination of the special microscopic states to characterized macroscopic physical property in equilibrium state.   }
\end{abstract}


\maketitle

\section{Introduction}
For classical many-body system where internal energy is the sum of kinetic and potential energy, physical quantity (especially, dynamical variables) in equilibrium state can be obtained through thermodynamic  average (the so-called canonical average), which includes summation taken over all microscopic states at provided composition on phase space. 
Since number of possible microscopic states astronominally increases with increase of system size, direct evaluation macroscopic physical property from the thermodynamic average is far from practical.  
To avoid such problem, several theoretical approaches have been amply developed to effectively sample important microscopic states to estimate dynamical variables, including Metropolis algorism, entropic sampling and Wang-Landau sampling.\cite{mc1, mc2, mc3, wl} 
Whereas physical quantities in equilibrium state can be reasonablly predicted by the existing theoretical approaches, a set of microscopic state to dominantly characterize equilibrium properties is generally unknown \textit{a priori}, without providing information about many-body interactions or temperature. This is a natural outcome because probability to find a chosen microscopic state $i$ is proportional to Boltzmann factor, $\exp\left(-\beta E_i\right)$. 

Despite these facts, we recently find a special set of microscopic structure (called "projection state: PS") constructed independently of information about many-body interaction and temperature, where their physical quantity can well characterize equilibrium properties including internal energy and macroscopic structure.\cite{em1,em2} This finding relies on the fact that configurational density of states (CDOS) before applying many-body interaction to the system is well-characterized by multidimensional gaussian when system size gets large.\cite{em3,em4} In our previous study, although condition of structure for PS is explicitly provided, analytical expression of the structure of PS on practical lattice remains unclear: The PS has been constructed based on the numerical simulation with special random sampling on configuration space. Therefore, in order to overcome such problem, we should clarify the quantitative landscape of CDOS for large systems. So far, it is not quantitatively clear why even one-dimensional CDOS along chosen coordination of pair correlation can be characterized by a gaussian-like distribution for typical periodic lattice. 

In the present study, we provide analytical expression of one-dimensional CDOS along selected pair figure on equiatomic binary system, by clarifying system-size dependence of any given moment of the CDOS. We demonstrate the validity of the derived expression by comparing the moment of CDOS obtained by numerical simulation for representative periodic lattices.

\section{Derivation and Applications}
Recently, we reveal analytical expression for composition dependence of 2nd order moment of CDOS for pair correlation on provided lattice. 
In that study, we employ generalized Ising model\cite{ce} (GIM) to quantitatively describe microscopic structure (i.e., atomic arrangement) on lattice, whose occupation at lattice site $i$ by A (B) element is given by the so-called spin variable, $\sigma_i = +1\,(-1)$. 
Briefly, correlation for microscopic structure $\vec{\sigma}$ along chosen pair $m$ in GIM is given by 
\begin{eqnarray}
\label{eq:pair}
\xi_m\left( \vec{\sigma} \right) = \Braket{\sigma_i \sigma_k}_{m,\textrm{lattice}},
\end{eqnarray}
where $\Braket{\cdot}_{m,\textrm{lattice}}$ denotes taking linear average over all symmetry-equivalent pair to $m$ on given lattice. 
With these preparations, we here extend our previous derivation for 2nd order moment of CDOS to any higher-order moment. In the same way to our previous approach, we start from rewriting pair correlation of Eq.~(\ref{eq:pair}) as
\begin{eqnarray}
\label{eq:xi}
\xi_m\left( \vec{\sigma} \right) = \left( 2D_mN \right)^{-1} \sum_{i,k} g_m\left( i,k \right) \sigma_i\left( \vec{\sigma} \right) \sigma_k\left( \vec{\sigma} \right),
\end{eqnarray}
where $D_m$ and $N$ denotes number of pair $m$ per site and number of lattice points in the system, respectively, and summation is taken over all lattice points. $g_m\left( i,k \right)$ takes 1 (0) if site $i$ and $k$ forms pair $m$ (for otherwise). 
Based on Eq.~(\ref{eq:xi}), $r$-th order moment of the CDOS along pair $m$ can be given by
\begin{widetext}
\begin{eqnarray}
\label{eq:mu}
\mu_r^{\left(m\right)} = \Braket{\xi_m^r}_{\vec{\sigma}} = \frac{1}{\left( 2D_mN \right)^r} \sum_{p_1,p_2}\sum_{p_3,p_4}\cdots\sum_{p_{2r-1},p_{2r}} g_m\left( p_1,p_2 \right) \cdots g_m\left( p_{2r-1}, p_{2r} \right) \Braket{\sigma_{p_1}\sigma_{p_2}\cdots\sigma_{p_{2r}}}_{\vec{\sigma}},
\end{eqnarray}
\end{widetext}
where $\Braket{\cdot}_{\vec{\sigma}}$ represents taking linear average over possible microscopic structure on given lattice.

\begin{figure}[h]
\begin{center}
\includegraphics[width=0.88\linewidth]{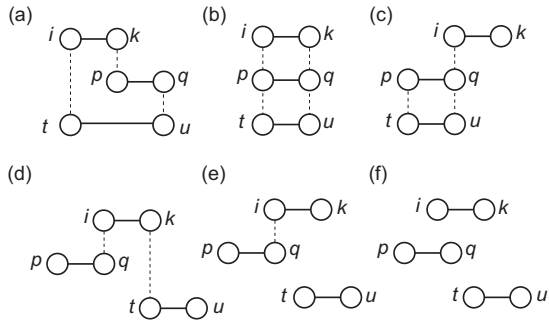}
\caption{ Figure desciption for possible combination of lattice points for third-order moment, appears in summation of Eq.~\ref{eq:mu}. Lattice points connected by vertical dotted lines corresponds to the same one.  }
\label{fig:mom3}
\end{center}
\end{figure}

We have shown that at equiatomic composition, linear average of four spin product $\Braket{\sigma_{p_1}\sigma_{p_2}\sigma_{p_3}\sigma_{p_4}}_{\vec{\sigma}}$ can be treated by taking product over independently occupied spin variables depending only on composition, i.e., $x=0.5$: Based on the idea, we have successfully provide composition dependence of  2nd order moment of CDOS for given pair. 
Not only in terms of considering the landscape of CDOS, but also of statistical independence of CDOS, treating multisite spin product as product of independent-occupation: Our previous study showed that density of eigenvalues for covariance matrix of practical CDOS at equiatomic composition become numerically identical to that obtained from random matrix with gaussian orthogonal ensemble, where each element of the matrix \textit{independently} takes normal random numbers, which directly means that for large systems, global landscape of CDOS become close to the DOS with independently-occupied random states.
When we extend the idea to higher-order moment, non-zero contribution to $\mu_r$ from $2r$-spin product, $\Braket{\sigma_{p_1}\sigma_{p_2}\cdots\sigma_{p_{2r}}}_{\vec{\sigma}}$, should always consists of even-times spin product for all constituent lattice points included in a set of $\left\{ p_1,p_2,\cdots,p_{2r} \right\}$, which naturally comes from symmetric definition of spin variable $\sigma=\pm 1$. 
For instance, when we estimate third order moment, six-spin product should be considered. Figure~\ref{fig:mom3} shows the possible combination of six-spin products $\Braket{\sigma_i \sigma_k \sigma_p \sigma_q \sigma_t \sigma_u}$, where lattice points connected with vertical broken line denotes that they corresponds to the same lattice point: e.g., in Fig.~\ref{fig:mom3} (a), $i=t$, $k=p$ and $q=u$, and in (b), $i=p=t$ and $k=q=u$. Among the six combinations, only (a) has non-zero contribution to $\mu_3$, since other five contains at least one odd-times spin product in the same lattice point. For instance, (c) has one even-time product at $p$ and $t$, while it has two odd-time products at $i$, $q$ $u$ and $k$. 
With this approach, we can significantly decrease the number of terms considered in Eq.~(\ref{eq:mu}). 

\begin{figure}[h]
\begin{center}
\includegraphics[width=0.8\linewidth]{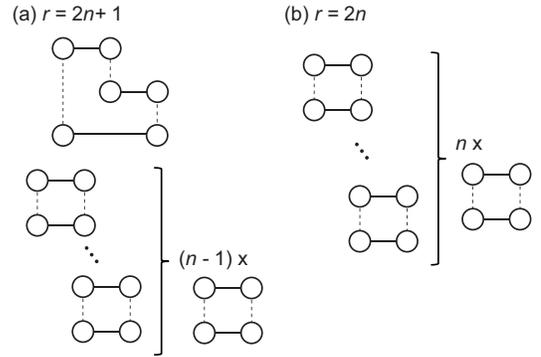}
\caption{ Figure description for maximum contribution (in terms of power of system size, $N$) to odd- (left) and even-order (right) moments. }
\label{fig:momr}
\end{center}
\end{figure}

However, when order of moment $r$ goes infinity, the number of terms with non-zero contribution to $\mu_r$ diverges. Therefore, additional strategy should be required to quantitatively determine $\mu_r$ for any given order. 
When we consider thermodynamic limit of $N\to\infty$, terms containing maximum power of $N$ in the summation $\sum_{p_1,p_2}\sum_{p_3,p_4}\cdots\sum_{p_{2r-1},p_{2r}}$ only contributes to $\mu_r$, since maximum power of $N$ in the summation for $r$-order moment is always less than $r$. 
Based on the idea in Fig.~\ref{fig:mom3}, such maximum contributions can be straightforwardly illustrated, for even- and odd-order moment as shown in Fig.~\ref{fig:momr}.
As shown, maximum contributions to each moment should be composed of $\left(n-1\right)$ two-pairs whose individual lattice point is the same (connected by dashed lines), and a single three-pairs (for odd moment) or a single two-pairs (for even moment). It is obvious that any other combinations of pairs always results in contribution of lower power of $N$ to the moment. Note that open circles that do not connected by dashed lines within the figure always correspond to different lattice point in the system. 
\begin{figure*}
\begin{center}
\includegraphics[width=0.62\linewidth]
{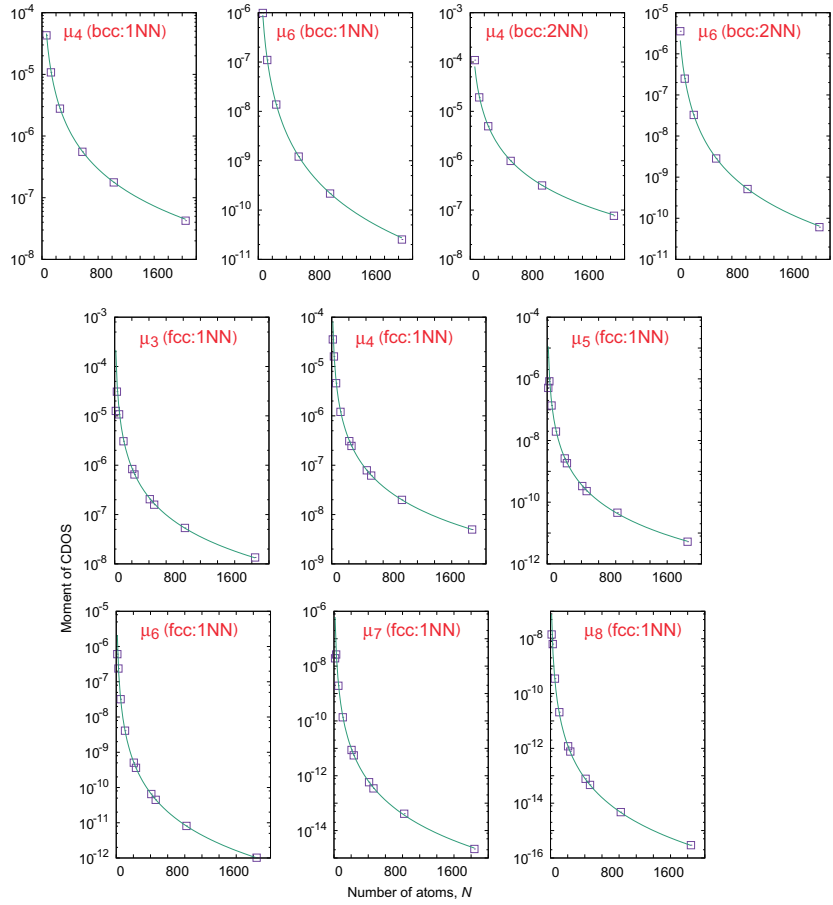}
\caption{ Moments of CDOS for 1NN and 2NN pair on fcc and bcc lattice, as a function of number of atoms in the system, $N$. Solid curves are obtained by the present expression of Eq.~(\ref{eq:muf}), and open squared denote numerical simulation. }
\label{fig:mom}
\end{center}
\end{figure*}
Therefore, when the system size $N$ gets large, moments for CDOS can be explicitly given by
\begin{eqnarray}
\label{eq:mm}
\mu_{2\alpha} &=& \mu_2^\alpha\cdot \left(2\alpha -1\right)!! \nonumber \\
\mu_{2\alpha + 1} &=& \mu_3\cdot \mu_2^{\left(\alpha -1\right)} \cdot {}_{2\alpha+1}\mathrm{C}_3 \cdot \left(2\alpha -3\right)!!,
\end{eqnarray}
where $\mu_2^\alpha$ and $ \mu_3\cdot \mu_2^{\left(\alpha -1\right)}$ for even- and odd-order moment in the equation respectively comes from the maximum contribution of Fig.~\ref{fig:momr}, and the rest terms of $\left(2\alpha -1\right)!!$ and ${}_{2\alpha+1}\mathrm{C}_3 \cdot \left(2\alpha -3\right)!!$ corresponds to the number of possible permutations to assign constituent pairs to the figure for maximum contribution of Fig.~\ref{fig:momr}. 
It has been shown\cite{sqs,cdos} that $\mu_2 = \left(D_mN\right)^{-1}$ at equiatomic composition. Thus, we here should determine the rest unknown term, $\mu_3$, as a function of $D_m$ and $N$. 
$\mu_3$ for $m$-th pair can be expressed as 
\begin{widetext}
\begin{eqnarray}
\mu_3^{\left(m\right)} = \frac{1}{\left( 2D_mN \right)^3} \sum_{i,k}\sum_{p,q}\sum_{t,u} g_m\left( i,k \right)  g_m\left( p,q \right) g_m\left( t,u \right)\Braket{\sigma_i\sigma_k\sigma_p\sigma_q\sigma_t\sigma_q}_{\vec{\sigma}}.
\end{eqnarray}
\end{widetext}
From Figs.~\ref{fig:mom3} and~\ref{fig:momr}, non-zero, maximum contribution to $\mu_3$ corresponds to the Fig.~\ref{fig:mom3} (a), thereby
\begin{eqnarray}
\label{eq:mu3}
\mu_3^{\left(m\right)} = \frac{ 2D_mN\cdot 4 \cdot 2 \cdot M_m   }{\left( 2D_mN \right)^3} = \frac{2 M_m}{\left( D_mN \right)^2},
\end{eqnarray}
where in the numerator of the first equation, $2D_mN$ corresponds to the number of ways to choose the first pair among three (e.g., $i-k$ pair), $4\cdot 2$ to the possible permutation of lattice points for the rest two pairs in the considered figure (i.e., Fig.~\ref{fig:mom3} (a)), and $M_m$ denotes the number of triples consisting of three $m$-th pairs where one of the three pair is kept fixed. 
Using Eqs.~(\ref{eq:mm}) and~(\ref{eq:mu3}), we can get final expression for the moments of CDOS for $m$-th pair, namely
\begin{eqnarray}
\label{eq:muf}
\mu_{2\alpha}^{\left(m\right)} &=& \frac{\left( 2\alpha-1\right)!!}{\left(D_mN\right)^\alpha} \nonumber \\
\mu_{2\alpha + 1}^{\left(m\right)} &=& \frac{2M_m\cdot {}_{2\alpha+1}\textrm{C}_3\cdot\left(2\alpha-3\right)!!}{\left(D_mN\right)^{\alpha+1}}  .
\end{eqnarray}
From Eq.~(\ref{eq:muf}), we can provide relationships between different order of moments by the following reccurence formula:
\begin{eqnarray}
\label{eq:rec}
\mu_{2\alpha}^{\left(m\right)} &=& \mu_{2\alpha-2}^{\left(m\right)} \cdot \left(2\alpha-1\right)\cdot\mu_{2}^{\left(m\right)}  \nonumber \\
\mu_{2\alpha+1}^{\left(m\right)}  &=& \mu_{2\alpha-1}^{\left(m\right)} \cdot \frac{\alpha\left(2\alpha+1\right)}{\left(\alpha-1\right)}\cdot \mu_{2}^{\left(m\right)}. 
\end{eqnarray}
It is now clear that the first equation of Eq.~(\ref{eq:rec}) is identical to the relationship of even-order moment for single-variate gaussian. Therefore, when all the odd-order moments are zero for chosen pair $m$, it is reasonable that the corresponding CDOS for large systems can be characterized by the gaussian. 
Such condition is satisfied when $M_m$ takes zero, i.e., there is no triplet consisting of three symmetry-equivalent $m$-th pairs on given lattice. For instance, 1NN pair on fcc lattice have $M_m=4$, while 2NN have $M_m=0$.

Finally, in order to demonstrate the validity of the derived analytical expression of the moments in Eq.~(\ref{eq:muf}), we perform Monte Carlo (MC) simulation for equiatomic binary system along 1NN on fcc and 1NN and 2NN pair on bcc lattices, where possible microscopic structures are uniformely sampled without any statistical weight.  Based on the simulation, we estimate from third- to eight-order moment as a function of number of lattice points in the system, $N$. The results are compared with the derived expression of Eq.~(\ref{eq:muf}), as shown in Fig.~\ref{fig:mom}. Note that pairs that have $M_m=0$ are excluded in Fig.~\ref{fig:mom}, since theorical prediction always show $N$-independence, while numerical simulation should always exhibit $N$ dependence. 
We can clearly see the excellent agreement of $N$-dependence of all the moments shown, indicating that the present theoretical approach can reasonably capture the landscape of CDOS in terms of simple geometric information of coordination number, number of specially-selected triplets and of the system size. 

\section{Conclusions}
For classical discrete systems at equiatomic composition, we propose analytical expression of any-order moments of configurational density of states (CDOS) for non-interacting system. 
Validity of the derived expression is demonstrated by comparing the moments obtained by numerical simulation as a function of the system size. 
We confirm that for large systems, landscape of CDOS can be reasonablly characterized by simple geometric information such as coordination number, number of specially-selected triplets and of the system size.

\section*{Acknowledgement}
This work was supported by a Grant-in-Aid for Scientific Research (16K06704) from the MEXT of Japan, Research Grant from Hitachi Metals$\cdot$Materials Science Foundation, and Advanced Low Carbon Technology Research and Development Program of the Japan Science and Technology Agency (JST).


\begin{thebibliography}{9}
\bibitem{mc1} N. Metropolis, A. W. Rosenbluth, M. N. Rosenbluth, A. H. Tellerand, and E. Teller, J. Chem. Phys. \textbf{21}, 1087 (1953).
\bibitem{mc2} A. M. Ferrenberg and R. H. Swendsen, Phys. Rev. Lett. \textbf{63}, 1195 (1989). 
\bibitem{mc3} J. Lee, Phys. Rev. Lett. 71, 211 (1993).
\bibitem{wl} F. Wang and D.P. Landau, Phys. Rev. Lett. \textbf{86}, 2050 (2001).
\bibitem{em1} K. Yuge, J. Phys. Soc. Jpn.  \textbf{84}, 084801 (2015).
\bibitem{em2} K. Yuge, J. Phys. Soc. Jpn. \textbf{85}, 024802  (2016).
\bibitem{em3} K. Yuge, T. Kishimoto and K. Takeuchi, Trans. Mat. Res. Soc. Jpn. \textbf{41}, 213 (2016).
\bibitem{em4} T. Taikei, T. Kishimoto, K. Takeuchi and K. Yuge, J. Phys. Soc. Jpn. (submitted).
\bibitem{ce} J.M. Sanchez, F. Ducastelle, and D. Gratias, Physica A \textbf{128}, 334 (1984).
\bibitem{sqs} S.-H. Wei, L. G. Ferreira, J. E. Bernard, and A. Zunger, Phys. Rev. B \textbf{42}, 9622 (1990).
\bibitem{cdos} K. Yuge, T. Taikei and K. Takeuchi, Trans. Mat. Res. Soc. Jpn. (in press).
\end{thebibliography}
\end{document}